\documentclass{appolb}
\usepackage{graphicx}

\begin{document}
\title{Beauty production in heavy-ion collisions with ALICE at the LHC%
\thanks{Presented at Quark Matter 2022, Krakow (Poland).}%
}
\author{ ${\rm Xinye}$~${\rm Peng}^{1,2}$ on behalf of the ALICE collaboration
\address{ $^{1}$China University of Geosciences (Wuhan), China}
\address{ $^{2}$Central China Normal University, China}
}
\maketitle
\begin{abstract}
In this contribution, the final measurements of the centrality dependence of the nuclear modification factor ($R_{\rm AA}$) of non-prompt $\rm D^0$ in Pb--Pb collisions at $\sqrt{s_{\scriptscriptstyle \rm NN}}$ = 5.02 TeV will be presented. These measurements provide important constraints to the mass dependence of in-medium energy loss and hadronisation of the beauty quark. The $p_{\rm T}$-integrated non-prompt $\rm D^0$ $R_{\rm AA}$ will be presented for the first time and will be compared to the prompt $\rm D^0$ one. This comparison will shed light on possible different shadowing effects between charm and beauty quarks. In addition, the first measurements of non-prompt $\rm D_s^+$ production in central and semi-central Pb--Pb collisions at $\sqrt{s_{\scriptscriptstyle \rm NN}}$ = 5.02 TeV  will be discussed. The non-prompt $\rm D_s^+$ measurements provide additional information on the hadronisation of beauty quarks and the production yield of $\rm B_s^{0}$ mesons. Finally, the first measurement of non-prompt D-meson elliptic flow in Pb--Pb collisions at $\sqrt{s_{\scriptscriptstyle \rm NN}}$ = 5.02 TeV will also be discussed. These measurements can constrain the degree of thermalisation of beauty quarks in the hot and dense QCD medium.
\end{abstract}
  
\section{Introduction}
Heavy quarks (charm and beauty) are produced in hard-scattering processes over short time scales compared to the quark--gluon plasma (QGP). They probe the whole system evolution interacting with the medium constituents. In particular, because of the larger mass, beauty quarks are expected to lose less energy~\cite{Zhang:2003wk,Djordjevic:2003zk} and diffuse less than the charm quarks~\cite{Moore:2004tg,Petreczky:2005nh}. Therefore, the comparison between charm and beauty nuclear modification factor ($R_{\rm AA}$) and elliptic flow ($v_2$) provides insight into the quark mass-dependent of energy loss, heavy-quark diffusion properties.

The D mesons from beauty-hadron decays (non-prompt D) are excellent probes for beauty properties currently. Existing data on the production of B mesons~\cite{CMS:2017uoy}, and J/$\psi$ from beauty decays~\cite{CMS:2017uuv,ALICE:2015nvt,Aaboud:2018quy} at midrapidity are limited by large uncertainties, while the broad correlation between the transverse momenta ($p_{\rm T}$) of the lepton and the parent beauty hadron reduces the effectiveness of beauty-decay lepton measurements~\cite{ALICE:2016uid,ATLAS:2021xtw}.

In these proceedings, the measurement of open-beauty production in Pb--Pb collisions at $\sqrt{s_{\scriptscriptstyle \rm NN}}$ = 5.02 TeV from ALICE at midrapidity is reported.

\section{D-meson reconstruction}

The non-prompt D mesons are reconstructed via their hadronic decay channels ${\rm D}^0 \to {\rm K}^-\pi^+$, and ${\rm D}_s^+ \to {\rm \pi}^+\phi \to {\rm \pi}^+ {\rm K}^+ {\rm K}^-$. A multi-classification BDT algorithm~\cite{Acharya:2021cqv}, trained using variables sensitive to the decay topology and particle identification, is utilised to simultaneously increase the non-prompt (${\rm b} \to {\rm D}$) fraction and suppress the combinatorial background. After extracting the yield via an invariant mass analysis, the ${\rm b} \to {\rm D}$ fraction is estimated by a $\chi^2$-minimisation approach based on variation of ML-based selections~\cite{Acharya:2021cqv}. The measurement of the non-prompt $\rm D^0$ $v_2$ is performed with the scalar-product (SP) method~\cite{Voloshin:2008dg}. In order to obtain non-prompt $\rm D^0$ $v_2$ ($v_2^{\rm non-prompt}$), a linear fit of $v_2^{\rm sum}$ is performed as a function of ${\rm b} \to {\rm D}$ fraction, and extrapolates to $f_{\rm non-prompt} = 1$, as described in~\cite{CMS:2020qul}.  

\section{Results}

\begin{figure}[h]
\begin{center}
\includegraphics[width=0.5\textwidth]{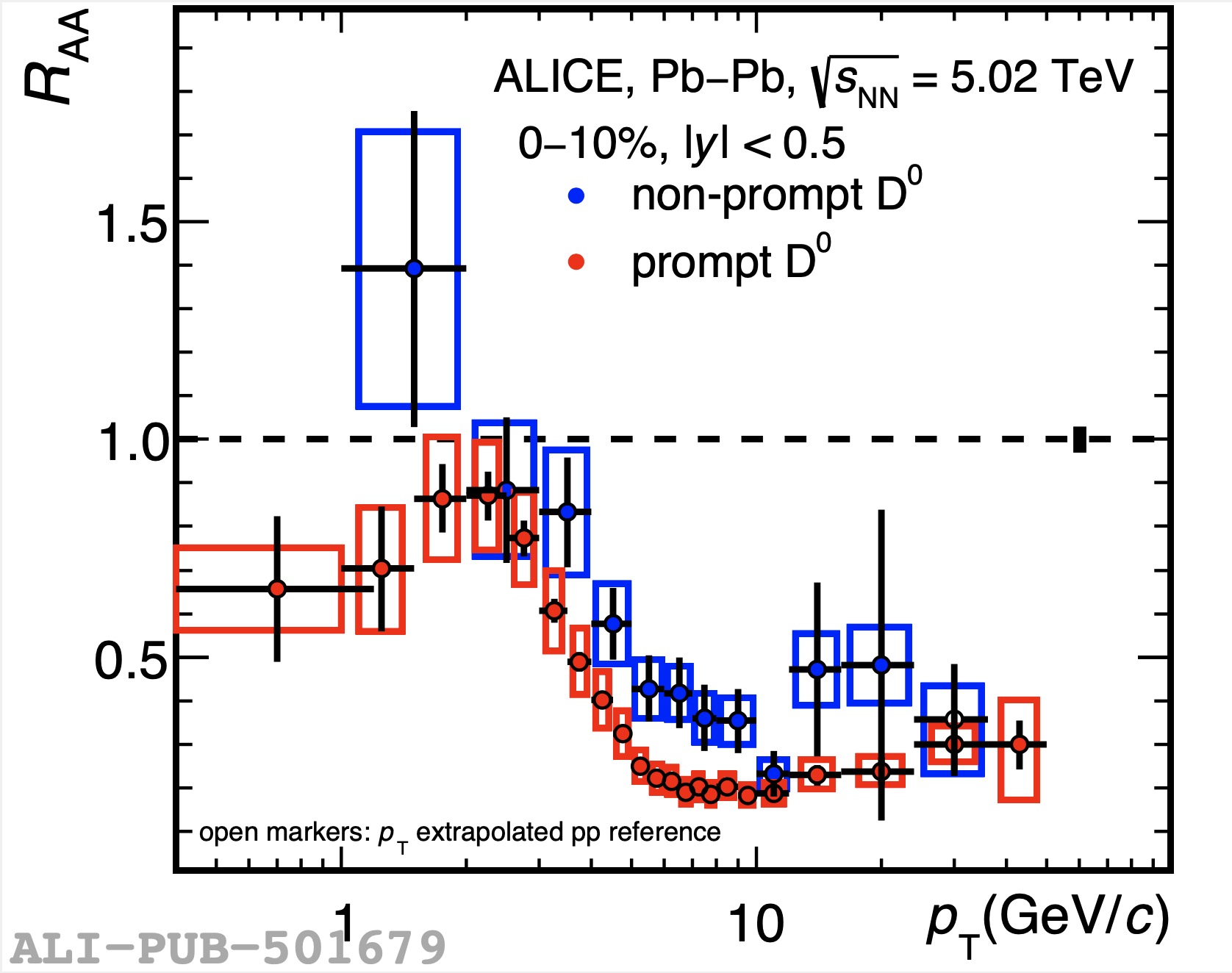}
\includegraphics[width=0.4\textwidth]{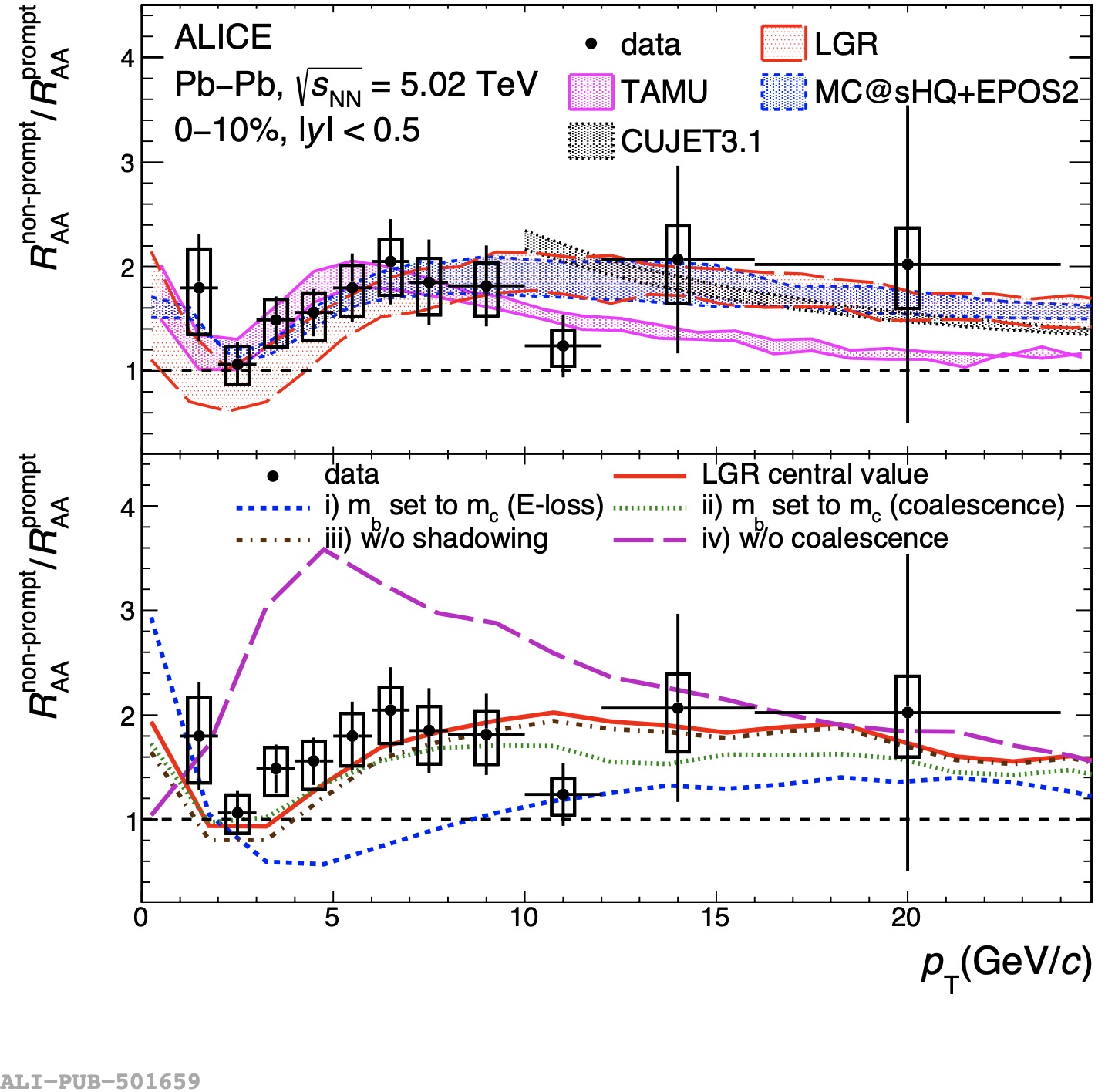}
\caption{Left panel: the $R_{\rm AA}$ of non-prompt $\rm D^0$ mesons~\cite{ALICE:2022tji} in the centrality classes 0--10\%, compared with the $R_{\rm AA}$ of prompt $\rm D^0$ mesons~\cite{ALICE:2021rxa}. Right panel: non-prompt to prompt $\rm D^0$-meson $R_{\rm AA}$ ratio
as a function of $p_{\rm T}$ in the 0--10\% central Pb--Pb collisions at $\sqrt{s_{\scriptscriptstyle \rm NN}}$ = 5.02 TeV, compared to  model predictions~\cite{Nahrgang:2013xaa,Li:2020umn,Li:2019lex,He:2014cla,Shi:2019nyp} (top), and to different modifications of LGR calculations~\cite{Li:2020umn,Li:2019lex} (bottom).}
\label{fig:p_np_raa}
\end{center}
\end{figure}

The left panel of Fig.~\ref{fig:p_np_raa} shows the $R_{\rm AA}$ of non-prompt $\rm D^0$ mesons~\cite{ALICE:2022tji} measured in the 0--10\% most central Pb--Pb collisions at $\sqrt{s_{\scriptscriptstyle \rm NN}}$ = 5.02 TeV. The result is compared with the prompt $\rm D^0$-meson $R_{\rm AA}$~\cite{ALICE:2021rxa}. The non-prompt $\rm D^0$-meson $R_{\rm AA}$ is significantly higher than the prompt $\rm D^0$ one for $p_{\rm T}>5$ GeV/$c$, indicating that non-prompt $\rm D^0$ mesons are less suppressed than prompt $\rm D^0$ mesons, and supporting the expectation that beauty quarks lose less energy than charm quarks because of their larger mass. An extrapolation of the measured spectrum is performed. The resulting non-prompt $\rm D^0$-meson $R_{\rm AA}$ for $p_{\rm T} > 0$ is $1.00\pm 0.10 \ ({\rm stat.})\pm 0.13 \ ({\rm syst.}){ }^{+0.08}_{-0.09} \ ({\rm extr.})\pm 0.02 \ ({\rm norm.})$~\cite{ALICE:2022tji} in the 0--10\% centrality class, which is compatible with unity within uncertainties, and larger than the prompt one~\cite{ALICE:2021rxa} within less than $1.5 \sigma$. 

The non-prompt to prompt $\rm D^0$-meson $R_{\rm AA}$ ratio as a function of $p_{\rm T}$ in the 0--10\% central Pb--Pb collisions at $\sqrt{s_{\scriptscriptstyle \rm NN}}$ = 5.02 TeV is presented in the right panel of Fig.~\ref{fig:p_np_raa}. The ratio can be described well by model predictions~\cite{Nahrgang:2013xaa,Li:2020umn,Li:2019lex,He:2014cla,Shi:2019nyp} that include collisional and radiative processes (top panel). In order to further understand the data pattern at low $p_{\rm T}$ and the enhancement with respect to unity at high $p_{\rm T}$ for the data, the ratio is compared with different LGR model~\cite{Li:2020umn,Li:2019lex} calculations (bottom panel), highlighting the role of coalescence, shadowing, and mass dependence of energy loss. The LGR model suggests that the “valley” structure at low $p_{\rm T}$ is mainly due to the formation of prompt D mesons via charm-quark coalescence (case iv), and the significant enhancement of the ratio at high $p_{\rm T}$ is interpreted as the effect of the mass dependence of the in-medium energy loss (case i).

\begin{figure}[h]
\begin{center}
\includegraphics[width=0.9\textwidth]{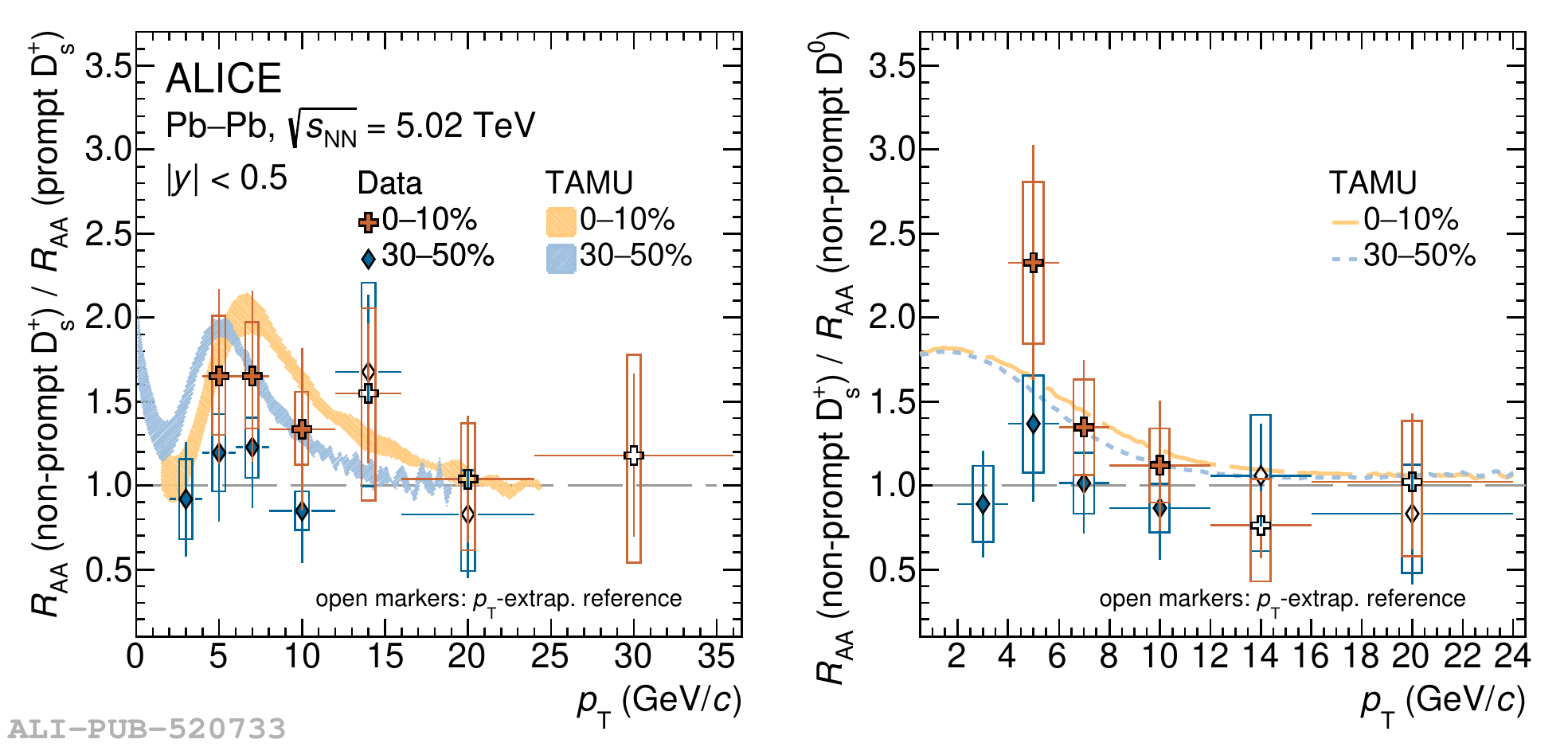}
\caption{The $R_{\rm AA}$ of non-prompt $\rm D_s^+$ mesons~\cite{ALICE:2022xrg} divided by the one of prompt $\rm D_s^+$ mesons~\cite{ALICE:2021kfc} (left panel) and non-prompt $\rm D^0$ mesons~\cite{ALICE:2022tji} (right panel) for the 0--10\% and 30--50\% centrality classes in Pb--Pb collisions at $\sqrt{s_{\scriptscriptstyle \rm NN}}$ = 5.02 TeV. The measurements are compared with TAMU model predictions~\cite{He:2014cla}.}
\label{fig:p_npDs_raa}
\end{center}
\end{figure}

The $R_{\rm AA}$ of non-prompt $\rm D_s^+$ mesons divided by that of prompt $\rm D_s^+$ mesons~\cite{ALICE:2021kfc} (left panel) and non-prompt $\rm D^0$ mesons~\cite{ALICE:2022tji} (right panel) are shown in Fig.~\ref{fig:p_npDs_raa}. In the 0--10\% centrality class, a hint that both ratios are larger than unity in the $4 < p_{\rm T} < 12$ GeV/$c$ interval is presented, suggesting an enhancement of non-prompt $\rm D_s^+$ in this interval. The trend can be explained by the interplay of mass dependence of energy loss and recombination in the QGP medium. The ratios are compatible with unity within uncertainties. The TAMU predictions~\cite{He:2014cla} qualitatively describe the results for central collisions. For semicentral collisions the TAMU model overestimates the $R_{\rm AA}$ ratio values.

\begin{figure}[h]
\begin{center}
\includegraphics[width=0.45\textwidth]{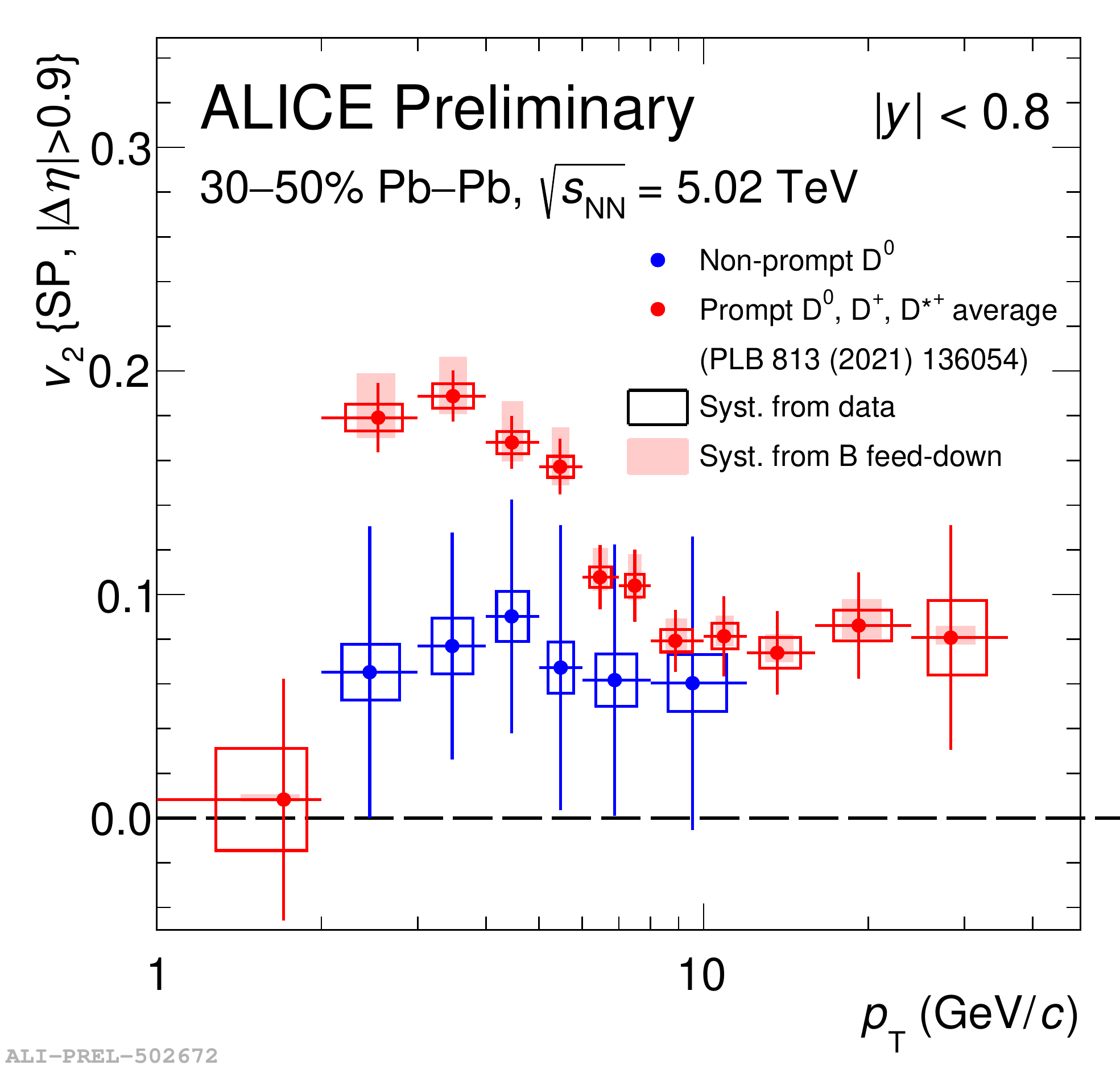}
\includegraphics[width=0.45\textwidth]{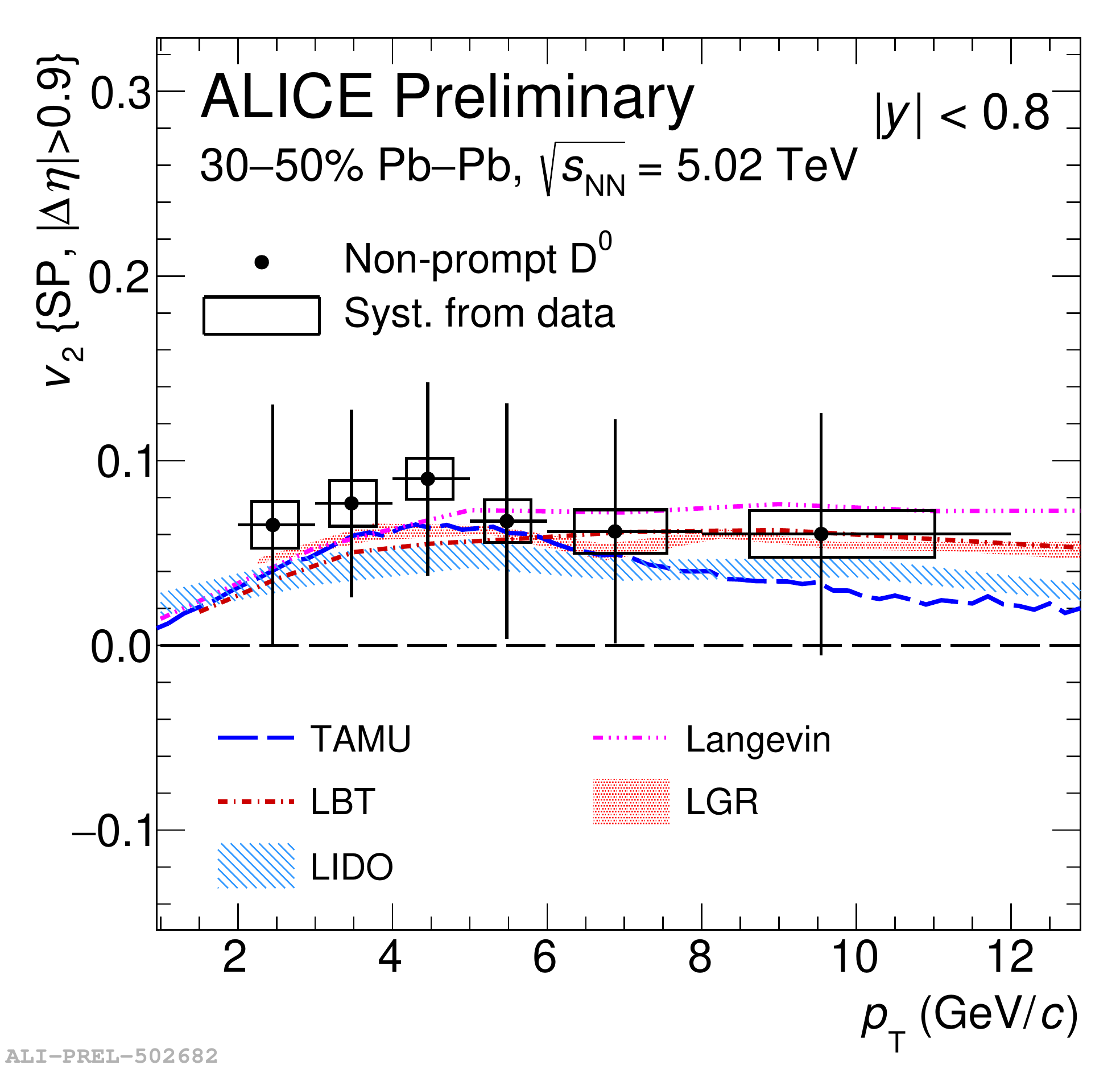}
\caption{The non-prompt $\rm D^0$ $v_2$ in Pb--Pb collisions at $\sqrt{s_{\scriptscriptstyle \rm NN}}$ = 5.02 TeV in the 30--50\% centrality class, compared to that of prompt average non-strange D mesons~\cite{ALICE:2020iug} (left panel), and to the model predictions~\cite{Li:2020umn,Li:2021xbd,Li:2020kax,He:2014cla,Ke:2018tsh,Ke:2018jem} (right panel).}
\label{fig:np_v2}
\end{center}
\end{figure}

Figure~\ref{fig:np_v2} shows the first measurement of non-prompt $\rm D^0$ $v_2$ in 30--50\% Pb--Pb collisions at $\sqrt{s_{\scriptscriptstyle \rm NN}}$ = 5.02 TeV, compared to the average $v_2$ of prompt D mesons~\cite{ALICE:2020iug} (left panel), and to model predictions~\cite{Li:2020umn,Li:2021xbd,Li:2020kax,He:2014cla,Ke:2018tsh,Ke:2018jem} (right panel). The non-prompt $\rm D^0$ $v_2$ is found to be positive with $2.7$ $\sigma$ significance for $2 < p_{\rm T} < 12$ GeV/$c$. The significance for the difference between non-prompt $\rm D^0$ $v_2$ and that of prompt average non-strange D mesons is $3.2$ $\sigma$ for $2 < p_{\rm T} < 8$ GeV/$c$, indicating a different degree of participation to collective motion between charm and beauty quarks. Theoretical predictions~\cite{Li:2020umn,Li:2021xbd,Li:2020kax,He:2014cla,Ke:2018tsh,Ke:2018jem} based on beauty-quark transport in an hydrodynamical expanding medium can fairly describe the data within uncertainties. Future measurements with higher accuracy will provide important constraints to the models, and allow for accurate extraction of the spatial diffusion coefficient with beauty quarks.

\section{Conclusion}
In this contribution, the most recent results on the production and azimuthal anisotropy of non-prompt D mesons, measured in Pb--Pb collisions at $\sqrt{s_{\scriptscriptstyle \rm NN}}$ = 5.02 TeV, were presented. The non-prompt $\rm D^0$ $R_{\rm AA}$  provides an essential constraint on the mass dependence of energy loss in the medium. The non-prompt $\rm D_s^+$ $R_{\rm AA}$ helps to further understand beauty-quark hadronisation in heavy-ion collisions. The first measurement of non-prompt $\rm D^0$-meson $v_2$ indicates a different degree of participation in the collective motion between charm and beauty quarks. With the LHC Run3, the upgraded ITS and the increased integrated luminosity will allow for more precise beauty-hadron measurements at midrapidity with ALICE.

\section{Acknowledgement}
This work was partly supported by the NSFC Key Grant 12061141008, the National key research and development program of China under Grant 2018YFE0104700 and 2018YFE0104800, the National Natural Science Foundation of China (Grant No. 12175085), and Fundamental research funds for the Central Universities (No. CCNU220N003).

\bibliographystyle{utphys}
\bibliography{qm22xinye}

\providecommand{\href}[2]{#2}\begingroup\raggedright\begin{thebibliography}{10}

\bibitem{Zhang:2003wk}
B.-W. Zhang, E.~Wang, and X.-N. Wang, ``{Heavy quark energy loss in nuclear
  medium}'', \href{http://dx.doi.org/10.1103/PhysRevLett.93.072301}{{\em
  Phys.Rev.Lett.} {\bfseries 93} (2004) 072301},
\href{http://arxiv.org/abs/nucl-th/0309040}{{\ttfamily arXiv:nucl-th/0309040
  [nucl-th]}}.

\bibitem{Djordjevic:2003zk}
M.~Djordjevic and M.~Gyulassy, ``{Heavy quark radiative energy loss in QCD
  matter}'', \href{http://dx.doi.org/10.1016/j.nuclphysa.2003.12.020}{{\em
  Nucl.Phys.} {\bfseries A733} (2004) 265--298},
\href{http://arxiv.org/abs/nucl-th/0310076}{{\ttfamily arXiv:nucl-th/0310076
  [nucl-th]}}.

\bibitem{Moore:2004tg}
G.~D. Moore and D.~Teaney, ``{How much do heavy quarks thermalize in a heavy
  ion collision?}'', \href{http://dx.doi.org/10.1103/PhysRevC.71.064904}{{\em
  Phys. Rev. C} {\bfseries 71} (2005) 064904},
  \href{http://arxiv.org/abs/hep-ph/0412346}{{\ttfamily arXiv:hep-ph/0412346}}.

\bibitem{Petreczky:2005nh}
P.~Petreczky and D.~Teaney, ``{Heavy quark diffusion from the lattice}'',
  \href{http://dx.doi.org/10.1103/PhysRevD.73.014508}{{\em Phys. Rev. D}
  {\bfseries 73} (2006) 014508},
  \href{http://arxiv.org/abs/hep-ph/0507318}{{\ttfamily arXiv:hep-ph/0507318}}.

\bibitem{CMS:2017uoy}
{\bfseries CMS} Collaboration, A.~M. Sirunyan {\em et~al.}, ``{Measurement of
  the ${B}^{\pm}$ Meson Nuclear Modification Factor in Pb-Pb Collisions at
  $\sqrt{{s}_{NN}} = 5.02$ TeV}'',
  \href{http://dx.doi.org/10.1103/PhysRevLett.119.152301}{{\em Phys. Rev.
  Lett.} {\bfseries 119} no.~15, (2017) 152301},
  \href{http://arxiv.org/abs/1705.04727}{{\ttfamily arXiv:1705.04727
  [hep-ex]}}.

\bibitem{CMS:2017uuv}
{\bfseries CMS} Collaboration, A.~M. Sirunyan {\em et~al.}, ``{Measurement of
  prompt and nonprompt charmonium suppression in Pb--Pb collisions at 5.02
  TeV}'', \href{http://dx.doi.org/10.1140/epjc/s10052-018-5950-6}{{\em Eur.
  Phys. J. C} {\bfseries 78} no.~6, (2018) 509},
  \href{http://arxiv.org/abs/1712.08959}{{\ttfamily arXiv:1712.08959
  [nucl-ex]}}.

\bibitem{ALICE:2015nvt}
{\bfseries ALICE} Collaboration, J.~Adam {\em et~al.}, ``{Inclusive, prompt and
  non-prompt J/$\psi$ production at mid-rapidity in Pb-Pb collisions at
  $\sqrt{s_{\rm NN}}$ = 2.76 TeV}'',
  \href{http://dx.doi.org/10.1007/JHEP07(2015)051}{{\em JHEP} {\bfseries 07}
  (2015) 051}, \href{http://arxiv.org/abs/1504.07151}{{\ttfamily
  arXiv:1504.07151 [nucl-ex]}}.

\bibitem{Aaboud:2018quy}
{\bfseries ATLAS} Collaboration, M.~Aaboud {\em et~al.}, ``{Prompt and
  non-prompt $J/\psi $ and $\psi (2\mathrm {S})$ suppression at high transverse
  momentum in $5.02~\mathrm {TeV}$ Pb+Pb collisions with the ATLAS
  experiment}'', \href{http://dx.doi.org/10.1140/epjc/s10052-018-6219-9}{{\em
  Eur. Phys. J. C} {\bfseries 78} no.~9, (2018) 762},
  \href{http://arxiv.org/abs/1805.04077}{{\ttfamily arXiv:1805.04077
  [nucl-ex]}}.

\bibitem{ALICE:2016uid}
{\bfseries ALICE} Collaboration, J.~Adam {\em et~al.}, ``{Measurement of
  electrons from beauty-hadron decays in p-Pb collisions at $
  \sqrt{s_{\mathrm{NN}}}=5.02 $ TeV and Pb-Pb collisions at $
  \sqrt{s_{\mathrm{NN}}}=2.76 $ TeV}'',
  \href{http://dx.doi.org/10.1007/JHEP07(2017)052}{{\em JHEP} {\bfseries 07}
  (2017) 052}, \href{http://arxiv.org/abs/1609.03898}{{\ttfamily
  arXiv:1609.03898 [nucl-ex]}}.

\bibitem{ATLAS:2021xtw}
{\bfseries ATLAS} Collaboration, G.~Aad {\em et~al.}, ``{Measurement of the
  nuclear modification factor for muons from charm and bottom hadrons in Pb+Pb
  collisions at 5.02 TeV with the ATLAS detector}'',
  \href{http://dx.doi.org/10.1016/j.physletb.2022.137077}{{\em Phys. Lett. B}
  {\bfseries 829} (2022) 137077},
  \href{http://arxiv.org/abs/2109.00411}{{\ttfamily arXiv:2109.00411
  [nucl-ex]}}.

\bibitem{Acharya:2021cqv}
{\bfseries ALICE} Collaboration, S.~Acharya {\em et~al.}, ``{Measurement of
  beauty and charm production in pp collisions at $\sqrt{s}=5.02$ TeV via
  non-prompt and prompt D mesons}'',
  \href{http://dx.doi.org/10.1007/JHEP05(2021)220}{{\em JHEP} {\bfseries 05}
  (2021) 220}, \href{http://arxiv.org/abs/2102.13601}{{\ttfamily
  arXiv:2102.13601 [nucl-ex]}}.

\bibitem{Voloshin:2008dg}
S.~A. Voloshin, A.~M. Poskanzer, and R.~Snellings, ``{Collective phenomena in
  non-central nuclear collisions}'',
  \href{http://dx.doi.org/10.1007/978-3-642-01539-7_10}{{\em Landolt-Bornstein}
  {\bfseries 23} (2010) 293--333},
  \href{http://arxiv.org/abs/0809.2949}{{\ttfamily arXiv:0809.2949 [nucl-ex]}}.

\bibitem{CMS:2020qul}
{\bfseries CMS} Collaboration, A.~M. Sirunyan {\em et~al.}, ``{Studies of charm
  and beauty hadron long-range correlations in pp and pPb collisions at LHC
  energies}'', \href{http://dx.doi.org/10.1016/j.physletb.2020.136036}{{\em
  Phys. Lett. B} {\bfseries 813} (2021) 136036},
  \href{http://arxiv.org/abs/2009.07065}{{\ttfamily arXiv:2009.07065
  [hep-ex]}}.

\bibitem{ALICE:2022tji}
{\bfseries ALICE} Collaboration, S.~Acharya {\em et~al.}, ``{Measurement of
  beauty production via non-prompt ${\rm D}^{0}$ mesons in Pb-Pb collisions at
  $\sqrt{s_{\rm NN}}$ = 5.02 TeV}'',
  \href{http://arxiv.org/abs/2202.00815}{{\ttfamily arXiv:2202.00815
  [nucl-ex]}}.

\bibitem{ALICE:2021rxa}
{\bfseries ALICE} Collaboration, S.~Acharya {\em et~al.}, ``{Prompt D$^{0}$,
  D$^{+}$, and D$^{*+}$ production in Pb\textendash{}Pb collisions at $
  \sqrt{s_{\mathrm{NN}}} $ = 5.02 TeV}'',
  \href{http://dx.doi.org/10.1007/JHEP01(2022)174}{{\em JHEP} {\bfseries 01}
  (2022) 174}, \href{http://arxiv.org/abs/2110.09420}{{\ttfamily
  arXiv:2110.09420 [nucl-ex]}}.

\bibitem{Nahrgang:2013xaa}
M.~Nahrgang, J.~Aichelin, P.~B. Gossiaux, and K.~Werner, ``{Influence of
  hadronic bound states above $T_c$ on heavy-quark observables in Pb + Pb
  collisions at at the CERN Large Hadron Collider}'',
  \href{http://dx.doi.org/10.1103/PhysRevC.89.014905}{{\em Phys. Rev. C}
  {\bfseries 89} no.~1, (2014) 014905},
  \href{http://arxiv.org/abs/1305.6544}{{\ttfamily arXiv:1305.6544 [hep-ph]}}.

\bibitem{Li:2020umn}
S.~Li, W.~Xiong, and R.~Wan, ``{Relativistic Langevin dynamics: charm versus
  beauty}'', \href{http://dx.doi.org/10.1140/epjc/s10052-020-08708-y}{{\em Eur.
  Phys. J. C} {\bfseries 80} no.~12, (2020) 1113},
  \href{http://arxiv.org/abs/2012.02489}{{\ttfamily arXiv:2012.02489
  [hep-ph]}}.

\bibitem{Li:2019lex}
S.~Li and J.~Liao, ``{Data-driven extraction of heavy quark diffusion in
  quark-gluon plasma}'',
  \href{http://dx.doi.org/10.1140/epjc/s10052-020-8243-9}{{\em Eur. Phys. J. C}
  {\bfseries 80} no.~7, (2020) 671},
  \href{http://arxiv.org/abs/1912.08965}{{\ttfamily arXiv:1912.08965
  [hep-ph]}}.

\bibitem{He:2014cla}
M.~He, R.~J. Fries, and R.~Rapp, ``{Heavy Flavor at the Large Hadron Collider
  in a Strong Coupling Approach}'',
  \href{http://dx.doi.org/10.1016/j.physletb.2014.05.050}{{\em Phys. Lett. B}
  {\bfseries 735} (2014) 445--450},
  \href{http://arxiv.org/abs/1401.3817}{{\ttfamily arXiv:1401.3817 [nucl-th]}}.

\bibitem{Shi:2019nyp}
S.~Shi, J.~Liao, and M.~Gyulassy, ``{Global constraints from RHIC and LHC on
  transport properties of QCD fluids in CUJET/CIBJET framework}'',
  \href{http://dx.doi.org/10.1088/1674-1137/43/4/044101}{{\em Chin. Phys. C}
  {\bfseries 43} no.~4, (2019) 044101},
  \href{http://arxiv.org/abs/1808.05461}{{\ttfamily arXiv:1808.05461
  [hep-ph]}}.

\bibitem{ALICE:2022xrg}
{\bfseries ALICE} Collaboration, ``{Measurement of beauty-strange meson
  production in Pb$-$Pb collisions at $\sqrt{s_{\rm NN}} = 5.02$ TeV via
  non-prompt $\mathrm{D_s}^{+}$ mesons}'',
  \href{http://arxiv.org/abs/2204.10386}{{\ttfamily arXiv:2204.10386
  [nucl-ex]}}.

\bibitem{ALICE:2021kfc}
{\bfseries ALICE} Collaboration, S.~Acharya {\em et~al.}, ``{Measurement of
  prompt $D_s^+$-meson production and azimuthal anisotropy in Pb\textendash{}Pb
  collisions at $\sqrt {s_{NN}}$=5.02TeV}'',
  \href{http://dx.doi.org/10.1016/j.physletb.2022.136986}{{\em Phys. Lett. B}
  {\bfseries 827} (2022) 136986},
  \href{http://arxiv.org/abs/2110.10006}{{\ttfamily arXiv:2110.10006
  [nucl-ex]}}.

\bibitem{ALICE:2020iug}
{\bfseries ALICE} Collaboration, S.~Acharya {\em et~al.},
  ``{Transverse-momentum and event-shape dependence of D-meson flow harmonics
  in Pb\textendash{}Pb collisions at $\sqrt {s_{NN}}$ = 5.02 TeV}'',
  \href{http://dx.doi.org/10.1016/j.physletb.2020.136054}{{\em Phys. Lett. B}
  {\bfseries 813} (2021) 136054},
  \href{http://arxiv.org/abs/2005.11131}{{\ttfamily arXiv:2005.11131
  [nucl-ex]}}.

\bibitem{Li:2021xbd}
S.-Q. Li, W.-J. Xing, X.-Y. Wu, S.~Cao, and G.-Y. Qin, ``{Scaling behaviors of
  heavy flavor meson suppression and flow in different nuclear collision
  systems at the LHC}'',
  \href{http://dx.doi.org/10.1140/epjc/s10052-021-09833-y}{{\em Eur. Phys. J.
  C} {\bfseries 81} no.~11, (2021) 1035},
  \href{http://arxiv.org/abs/2108.06648}{{\ttfamily arXiv:2108.06648
  [hep-ph]}}.

\bibitem{Li:2020kax}
S.-Q. Li, W.-J. Xing, F.-L. Liu, S.~Cao, and G.-Y. Qin, ``{Heavy flavor
  quenching and flow: the roles of initial condition, pre-equilibrium
  evolution, and in-medium interaction}'',
  \href{http://dx.doi.org/10.1088/1674-1137/abadee}{{\em Chin. Phys. C}
  {\bfseries 44} no.~11, (2020) 114101},
  \href{http://arxiv.org/abs/2005.03330}{{\ttfamily arXiv:2005.03330
  [nucl-th]}}.

\bibitem{Ke:2018tsh}
W.~Ke, Y.~Xu, and S.~A. Bass, ``{Linearized Boltzmann-Langevin model for heavy
  quark transport in hot and dense QCD matter}'',
  \href{http://dx.doi.org/10.1103/PhysRevC.98.064901}{{\em Phys. Rev. C}
  {\bfseries 98} no.~6, (2018) 064901},
  \href{http://arxiv.org/abs/1806.08848}{{\ttfamily arXiv:1806.08848
  [nucl-th]}}.

\bibitem{Ke:2018jem}
W.~Ke, Y.~Xu, and S.~A. Bass, ``{Modified Boltzmann approach for modeling the
  splitting vertices induced by the hot QCD medium in the deep
  Landau-Pomeranchuk-Migdal region}'',
  \href{http://dx.doi.org/10.1103/PhysRevC.100.064911}{{\em Phys. Rev. C}
  {\bfseries 100} no.~6, (2019) 064911},
  \href{http://arxiv.org/abs/1810.08177}{{\ttfamily arXiv:1810.08177
  [nucl-th]}}.

\end{thebibliography}\endgroup


\providecommand{\href}[2]{#2}\begingroup\raggedright\endgroup

\end{document}